# A Dynamics and Stability Framework for Avian Jumping Take-off

**Ben Parslew\*, Girupakaran Sivalingam, William Crowther**

*Department of Mechanical, Aerospace and Civil Engineering, The University of Manchester, Manchester, M13 9PL, United Kingdom*

## Summary

Jumping take-off in birds is an explosive behaviour with the goal of providing a rapid transition from ground to airborne locomotion. An effective jump is predicated on the need to maintain dynamic stability through the acceleration phase. The present study concerns understanding how birds retain control of body attitude and trajectory during take-off. Cursory observation suggests that stability is achieved with relatively little cost. However, analysis of the problem shows that the stability margins during jumping are actually very small and that stability considerations play a significant role in selection of appropriate jumping kinematics.

We use theoretical models to understand stability in prehensile take-off (from a perch) and also in non-prehensile take-off (from the ground). The primary instability is tipping, defined as rotation of the centre of gravity about the ground contact point. Tipping occurs when the centre of pressure falls outside the functional foot. A contribution of the paper is the development of graphical tipping stability margins for both centre of gravity location and acceleration angle. We show that the nose-up angular acceleration extends stability bounds forward and is hence helpful in achieving shallow take-offs. The stability margins are used to interrogate simulated take-offs of real birds using published experimental kinematic data from a guinea fowl (ground take-off) and a diamond dove (perch take-off). For the guinea fowl the initial part of the jump is stable, however simulations exhibit a stuttering instability not observed experimentally that is probably due to absence of compliance in the idealised joints. The diamond dove model confirms that the foot provides an active torque reaction during take-off, extending the range of stable jump angles by around 45°.

## 1. Introduction

Take-off is the most energetically demanding phase of flight, where the highest accelerations are imposed on the body. For birds it is well-understood that a leg-driven jump is advantageous for accelerating into flight (1–3). The energetic cost of jumping using ground reaction of muscle forces is much less than the energetic cost of using aerodynamic reaction in flapping flight, and also the maximum available force is higher (4,5). Thus, jumping is both more efficient and more effective than flapping wings as a means of propulsion. As the jump can only be employed briefly at the start of flight the efficiency benefit is only impactful in short commutes by avoiding time spent in low-speed, high-power flight conditions (6,7). However, the effectiveness benefit of using legs to achieve high acceleration applies equally well to flights of all durations.

The challenge of accelerating to a high take-off speed is equivalent to achieving a large jump height, which has been examined using theoretical dynamics models of humans (8–11), other mammals (12,13), and insects (e.g. starting with (14)); these studies provide a valuable framework for understanding limitations on jumping performance with known biomechanical constraints, such as limits on muscle force or power generation. In practical applications the development of robotic jumping devices (summarised in (15)) shares the ambition of maximising jump height, with given technological constraints of actuator force/torque and power output: the "Mowgli" frog-like robot (16) is capable of jumping to over half its own body height driven by pneumatic actuators using basic open-loop control; the Galago-inspired jumping robot (17) used a specialized leg mechanism with a series-elastic actuator[1] and jumps over eight times its own height. In both engineered and biological jumping systems research has focussed on understanding how stored energy is converted to maximise kinetic energy at the point of take-off. But the success of a jump is also predicated on the need to sustain an acceptable degree of control over the body trajectory and attitude (18,19). This is particularly relevant to birds where a range of take-off trajectories are needed to suit different behaviours, and where the body attitude directly influences the flight mechanics. The way in which birds select and sustain trajectory and attitude during take-offs has not yet been addressed, and forms the context for the present work.

\*Author for correspondence (ben.parslew@manchester.ac.uk)

To address this problem we will build upon well established stability concepts from literature on legged locomotion. The pioneering work of (20) provides fundamental mathematical expressions for stability of a legged system, and defines 'body stability' as the tendency of the body to return to its equilibrium attitude and altitude following a disturbance; the same definition is used here. The term 'zero tipping-moment point' (ZMP) later emerged as a useful metric for assessing body stability (e.g. see review in (21)): the ZMP was originally defined as the point on the foot where the resultant force could be applied (to yield zero net moment) (22), and as such the ZMP should remain within the base of support for stability.

Shih (23) used the ZMP as a stability metric to design motion trajectories for a stable bipedal walking robot; the robot constrained the ZMP to within the base of support and was able to demonstrate stable walking with the centre of gravity within the base of support ('statically stable') and beyond the base of support ('dynamically stable'). Li et al (24) developed a method to derive ZMP from experimental measurements of force and torque onboard a bipedal robot; the aim was to enable future devices to use measured ZMP as an input for closed loop control, although the authors noted substantial differences between predicted and measured ZMP. Nonetheless this proposed method has been implemented in stable bipedal walking robots (25) and continuous hopping robots (26). The ZMP was used to predict stable motion trajectories, and was also measured using a force/torque sensor as part of the closed loop controller. In both cases the systems exhibited stable behaviour, with the ZMP providing a robust yet conceptually simple measure of stability. We therefore see potential in applying this metric to provide insight into avian jumping. More generally, we envisage such stability metrics could be employed to offer further insight into stability of other jumping animals; this includes insects, where some quantitative analysis has been from using experiments and simulations of jumping locusts (18,19). Here, we will conduct a formalised analysis of stability, using 'centre of pressure' (CoP) as a stability metric, which has been shown to be equivalent to ZMP (27).

For clarity, we distinguish here between the CoP stability metric and those used to assess stability of running birds by examining convergence of motion amplitude (28,29), which are less applicable to jumping take-offs. We also highlight the simplified stability metrics that have been developed using abstractions of human walking dynamics (30,31). These approaches provide a useful estimate of stability using inverted pendulum models, but are neglected here in favour of CoP which captures more detailed system dynamics, including rotational momentum.

---

[1] A series elastic actuator contains a compliant element between the actuator and load, and can increase the instantaneous power availability compared to the actuator alone

Thus, a primary objective of our study is to apply existing CoP-based metrics in the context of avian jumping take-off to address the topic of stability, which will complement previous work on jumping energetics. Within this approach we will propose new methods to visualise clearly the margins of stability throughout the extension phase of a jumping take-off. To achieve this we will develop an avian dynamic model. The outputs from our study aim to enhance understanding and interpretations of existing avian jumping kinematic data in literature.

## 2. Theory

## 2.1 Definitions of jumping system components

We define a generic jumping system to comprise three functional components: body, leg and foot (Figure 1). The 'body' refers to the mass group including the head and trunk; the leg is an extensible member that is able to do mechanical work through linear and rotational displacements of the body and foot; the functional foot provides a distributed mechanical interface between the end of the leg and ground; the term 'foot', chosen here for brevity, encompasses the front and rear digits in the context of avian morphology (32).

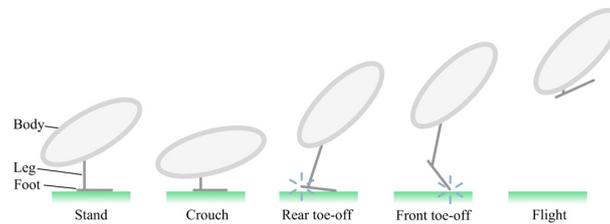

Figure 1: Functional components of body, leg and foot for a generic jumping system, with reference poses defining an avian jumping take-off. Blue spark icons indicate loss of contact with the ground.

The functional leg in a jumping system is characterised primarily by its stroke length, i.e. the maximum length it can extend, and the general force and power characteristics of the actuation system as a function of displacement and velocity. At the next level of detail (see section 3) we define the general kinematic arrangement of the mechanical components that comprise the leg and define its number of independent degrees of freedom. Lastly, for biological systems, we arrive at the detailed anatomy of the leg in terms of bones, muscles, tendons and other tissues. Our present work considers the leg down to the level of detail of kinematics only and does not comment directly on the physiology.

## 2.2 Take-off phases

Various levels of kinematic detail have been reported for jumping birds, from simple body trajectories (33,34) to reconstructions of joint angle time histories for the wing and leg (1,2). To frame the typical range of kinematic events observed in these jumps we define a series of different poses through which birds transition, Figure 1. *Stand* represents a nominal resting pose prior to initiation of a jump. To prepare for the jump the legs are retracted and the body rotated such that the body centre of gravity is lowered towards the ground. The *crouch* pose is chosen here as the pose of the body when the centre of gravity is at its lowest point (2). The principal mechanical work of the jump takes place during the acceleration from crouch to *front toe-off*, the point at which contact with the ground is lost and the ground reaction vanishes to zero. *Rear toe-off* is an intermediate event, which marks a change in the leg kinematics whereby the functional foot changes from being a static supporting element to being an additional link in the leg kinematic chain. With only the front-toe in contact with the ground the foot can no longer provide reaction torques and the system is at best neutrally stable[2].

Note that some animals perform a deliberate and sizeable preparatory hop prior to jumping (35–37), which is typically referred to as a stutter jump (38,39). This behaviour is not observed in avian take-off. However, we choose to maintain the use of the word 'stuttering' as a more general definition of any intermittent contact between foot and ground during the acceleration phase of the jump, prior to final toe-off.

## 2.3 Analytical models of toe-off and tipping stability

This section develops existing theory to provide simple analytical expressions and context for interrogation of the more complex multi-degree of freedom models presented later in section 2.4. The models considered here are: 1) a single degree of freedom model to understand the fundamentals of toe-off, including intermittent ground contact; 2) a planar dynamic model to understand the dynamics of tipping during take-off.

The single degree of freedom vertical jumping model (Figure 2) consists of a lumped mass body, $m$, a massless foot, and a vertically extensible leg of length $L$ with arbitrary extension kinematics. A similar but more involved analysis can be conducted for a leg with mass, but we show later that for avian jumping the leg mass makes little difference to the jumping dynamics.

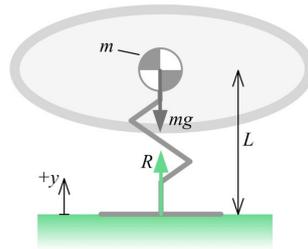

Figure 2 1D Toe-off model with vertically extensible leg; 1 translatory degree of freedom along the $y$-axis. Mass is concentrated at the body as indicated by the centre of gravity symbol.

The body centre of gravity is at a height, $y$, above the ground plane.

From dynamic equilibrium of forces the reaction $R$ at the foot is

$$R = mg + m\ddot{y}, \qquad 1$$

where $y$ is the height of the body centre of gravity above the ground plane, and the constant $g$ (magnitude of the gravitational acceleration vector) = 9.81ms$^{-2}$. Take-off will occur when $R$=0. The body acceleration condition for take-off is thus

$$\ddot{y} = -g = \ddot{L}, \qquad 2$$

where we distinguish between $\ddot{y}$ and $\ddot{L}$ so that the model is generalizable to conditions after toe-off. Equation 2 shows that for a massless leg system toe-off is governed purely by the vertical acceleration kinematics of the leg. If the foot has mass then the magnitude of deceleration of the body for take-off will be greater than 1$g$. This is intuitive as heavy feet mean that the force and hence body deceleration required for take-off is increased. The value in presenting equation 2 is that for a system with known

---

[2] A neutrally stable system does not tend to equilibrium, or deviate from it, following a disturbance

leg kinematics but without known ground reaction force (e.g. a jumping sequence recorded by video only) a prediction can be made about the time of toe-off purely from the system acceleration profile.

In a trivial sense, toe-off can happen during the stand posture by a sufficiently rapid contraction of the leg. This suggests a limit on the practical rate at which the bird can lower its body during the crouch phase. In a realistic case, toe-off will occur while the leg is still extending after the crouch phase at the point when the leg deceleration matches the condition in equation 2; this deceleration may be due to muscle actuation, or due to anatomical restrictions (ligaments) of joint rotation. This phenomenon is illustrated in Figure 3 and the supplementary videos S1, S2.

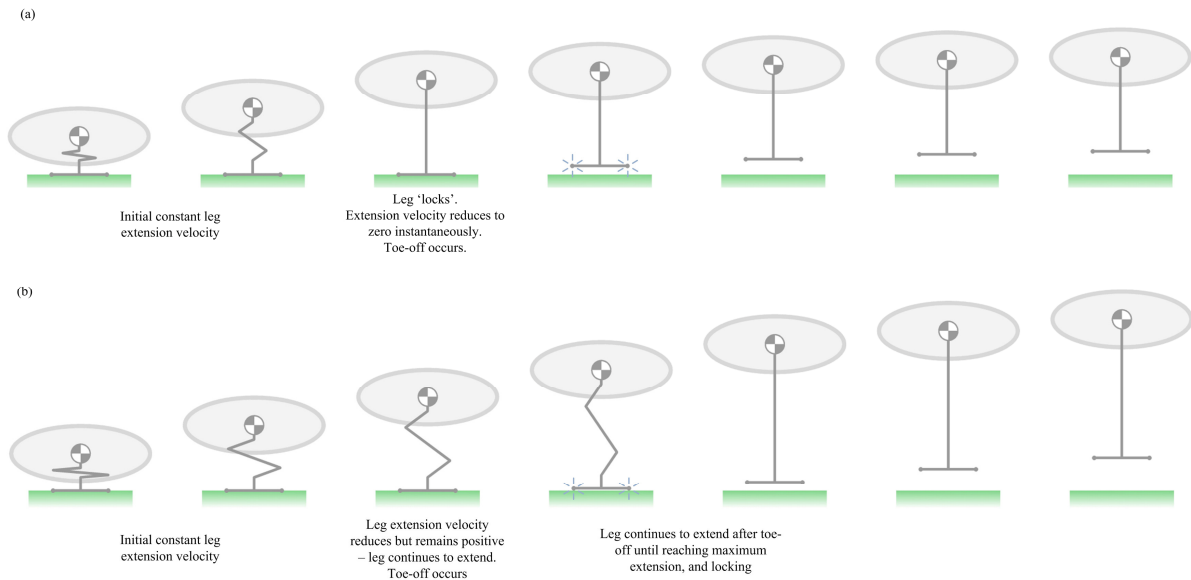

Figure 3 1D vertical jump illustrated using a body and a 'massless' leg.. The ziglag leg is an extensible element that can change length with any kinematic profile. In both examples toe-off occurs in the 3$^{rd}$ frame. (a) The leg extends at a constant velocity until it reaches maximum extension, at which point it locks instantaneously (negative infinite acceleration) causing toe-off (electronic supplementary material, video S1). (b) The leg extends with the same constant initial velocity as in (a), the velocity reduces (3rd frame) and remains positive, but as the acceleration is less than -$g$ toe-off occurs despite the fact that the leg isn't fully extended (electronic supplementary material, video S2).

Another simple, physically intuitive example of 1D jumping take-off is to replace the extensible leg by a linear spring, which applies a force proportional to the displacement from its natural length (supplementary material 1). This example is useful in showing that the body velocity has passed its peak value when toe-off occurs. With particular leg extension profiles it is possible for multiple toe-offs and flight phases to occur prior to the leg reaching its maximum extension: assuming a massless foot, if the body acceleration falls to -1$g$ toe-off occurs. But an increase in leg acceleration ($\ddot{L}$) would then cause the foot to regain contact with the ground. A subsequent toe-off will then occur when the body acceleration again reduces to -1$g$. These premature take-offs, also known as "stutters", have previously been avoided in spring-driven jumping robot designs in order to maximise conversion of stored elastic energy to kinetic energy (40). More recently it has been suggested that stutter jumps can achieve comparable jump heights to single jumps for robots, but with lower actuation power (39,40); these stutter jumps were inspired by the preliminary hops seen prior to take-off in some animals (35–37).

Extending the analysis to a 2D planar system allows us to consider a special case of toe-off, where the support polygon provided by the functional foot reduces to a single point, leading to tipping (rotation of the centre of gravity about the contact point of the foot with the substrate). Figure 4 shows a 2D planar jumping model for the purposes of illustrating tipping. Similar to Figure 2, this model has an extensible leg, but also includes two additional rotation degrees of freedom – one at the body-leg interface, and one at the leg-foot interface (open circles in Figure 4). The foot and leg are assumed to be massless. The body experiences a linear acceleration, $\ddot{\mathbf{r}}_B=[\ddot{x}_B, \ddot{y}_B, 0]$, and angular acceleration, $\dot{\boldsymbol{\omega}}_B$. These are reacted as an inertial force, $\mathbf{F}_R$, and inertial torque, $\mathbf{T}_R$ ($=[0,0,T_B]$, where $T_B$ is the torque applied to the body, positive anticlockwise). Contact points at the rear- and front-toe (grey-filled circles labelled "A" and "C", respectively, in Figure 4)[3] are underactuated unilateral joints (41) in that they can produce ground normal forces in one direction only and there is no control actuation at them (see section 3). The joints are bilateral in the horizontal direction in that friction can react forces in either direction, and slipping is impossible. The ground reaction forces at the rear- and front-toe are $\mathbf{R}_A$ and $\mathbf{R}_C$.

The net ground reaction force, $\mathbf{R}_P$, is illustrated in Figure 4 as acting at the CoP, $P$: the point on the ground plane through which $\mathbf{R}_P$ must pass in order for there to be zero net moment acting on the overall system. Equivalent to the ZMP (27), the CoP must be located within the convex hull of the foot or feet contact area(s) to maintain stability. In this 2D model, if the CoP passes ahead of the front-toe the system will tip forwards (clockwise), and if it passes behind the rear-toe the system will tip backwards (anticlockwise).

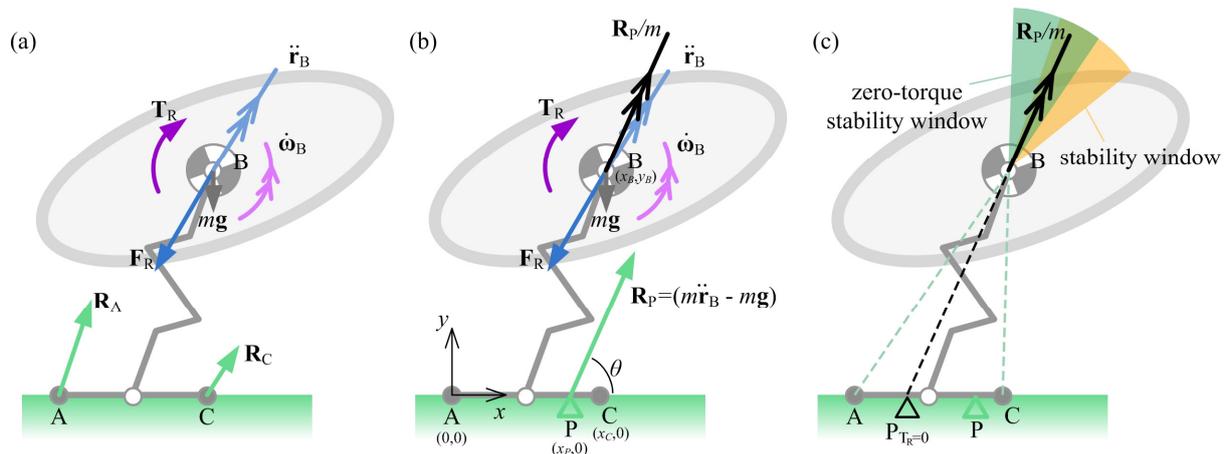

Figure 4 Diagram depicting the planar jumping system with body, extensible leg and foot with ground contact points at the front-and rear-toe (*A* and *C*, grey circles). Revolute joints are located at the body-leg and leg-foot interface (open circles). Body linear and angular acceleration are shown as vectors $\ddot{\mathbf{r}}_B$ and $\dot{\boldsymbol{\omega}}_B$. (a,b) The inertial force and torque are shown as vectors $\mathbf{F}_R$ and $\mathbf{T}_R$. Ground reaction forces at the front- and rear-toe are depicted as $\mathbf{R}_A$ and $\mathbf{R}_C$, and the net ground reaction force, $\mathbf{R}_P$, is equal to the vector sum of the body inertial and gravitational forces, and gives rise to the net acceleration $\mathbf{R}_P/m$. (b) The coordinate system is depicted for calculating the location of the centre of pressure, *P*, shown as a green triangle. The acceleration angle, $\theta$, is shown as the angle between the horizontal (*x*) axis and the ground reaction force vector. (c) The coloured sectors are stability windows within which the acceleration vector must point to avoid tipping. $P_{T_R=0}$ shows the hypothetical location of the centre of pressure for zero inertial torque ($\mathbf{T}_R=0$). The green sector is the case with zero angular acceleration (zero torque), the yellow is the case with positive, nose-up angular acceleration, and the overlapping region of these two sectors is illustrated as the dark green sector. The edges of the green sector align with the front and rear of the foot, as shown by the dotted lines. A similar (but less intuitive) graphical definition of the yellow stability window is that its edges would align with the front and rear of the foot if the foot were displaced backwards by a distance equal to the distance between $P_{T_R=0}$ and $P$; this has been omitted from the diagram for clarity.

---

[3] Any number of contact points could be used here with an equivalent result. In section 3 we develop a simulation model that applies ground contact forces at two fixed points on the foot as this is simpler than applying a single force at a moveable centre of pressure. To maintain continuity between models, we choose two contact points in this analytical model also.

The elevation angle of the net ground reaction force, termed here the 'acceleration angle', $\theta$, can be defined as

$$\theta = \tan^{-1}\left(\frac{\ddot{y}_B - g}{\ddot{x}_B}\right), \qquad (3)$$

and the horizontal displacement of the CoP from the origin at the rear toe can be written as:

$$x_P = x_B - \frac{y_B}{\tan\theta} + \frac{T_B}{\|\mathbf{R}_P\|\sin\theta} \qquad (4)$$

which is equivalent to the standard ZMP equation (see e.g. (42); full derivation shown in the supplementary materials). Equation 4 can be used to gather a broad understanding of how the applied force and torque affect the stability of the jump.

In the case of zero angular acceleration the reaction torque ($\mathbf{T_R}$) is zero and the CoP is located at the position $P_{TR=0}$ in Figure 4c. This position is the point on the ground plane which intersects a line drawn parallel to the net acceleration vector passing through the centre of mass (Figure 4c, black arrow). By definition, for the system to remain stable the CoP must lie between the rear- and front-toe, $0 < x_P < x_C$. So with zero angular acceleration for stability the acceleration vector must remain within the green sector shown in Figure 4c, termed here a 'stability window'. With positive (nose-up) angular acceleration the CoP shifts forwards to position $P$ and the stability window rotates clockwise – shown as the yellow sector in Figure 4c.

The forward stability bound is given by $x_P = x_C$ and the aft bound is given by $x_P = 0$. Substitution of these values in to equation 4 then allows us to define the forward and aft limits of the horizontal location of the centre of gravity, $x_B$, for stability under different acceleration conditions. The bounds are:

$$x_{B,aft} = \frac{y_B}{\tan\theta} - \frac{T_B}{\|\mathbf{R}_P\|\sin\theta}, \qquad (5)$$

$$x_{B,fore} = x_{B,aft} + x_C. \qquad (6)$$

There are three independent variables: the applied force magnitude, $\|\mathbf{R}_P\|$, its direction $\theta$, and the magnitude of the applied torque $T_B$. To simplify presentation we make the bounds dimensionless by dividing through by the length of the foot, $x_C$:

$$x'_{B,aft} = \frac{y_B'}{\tan\theta} - \frac{T_B'}{\sin\theta}, \qquad (7)$$

$$x'_{B,fore} = x'_{B,aft} + 1, \qquad (8)$$

where $T_B' = T_B/(\|\mathbf{R}_P\|x_C)$. Finally to close the problem we need to specify the value of $y_B'$ (the height of the centre of gravity relative to the length of the foot). As a simple test case, let us assume $y_B' = 1$. Figure 5 illustrates the stability bounds for jumps with acceleration angles from 0 to 90°, and for three different cases of applied torque. Consider the zero-torque (red) case first. If the acceleration vector is vertical (i) then the jump is stable if the horizontal location of the centre of gravity falls between the front- and rear-toe, as suggested by intuition. If the jump is in a forward direction (iv,v) then for stability the acceptable centre of gravity range is shifted forward, with shallower jumps requiring a bigger shift; (v) highlights that for acceleration angles of 45° or shallower the centre of gravity range for dynamic stability is outside of the range needed for static stability (i.e. outside of the foot region).

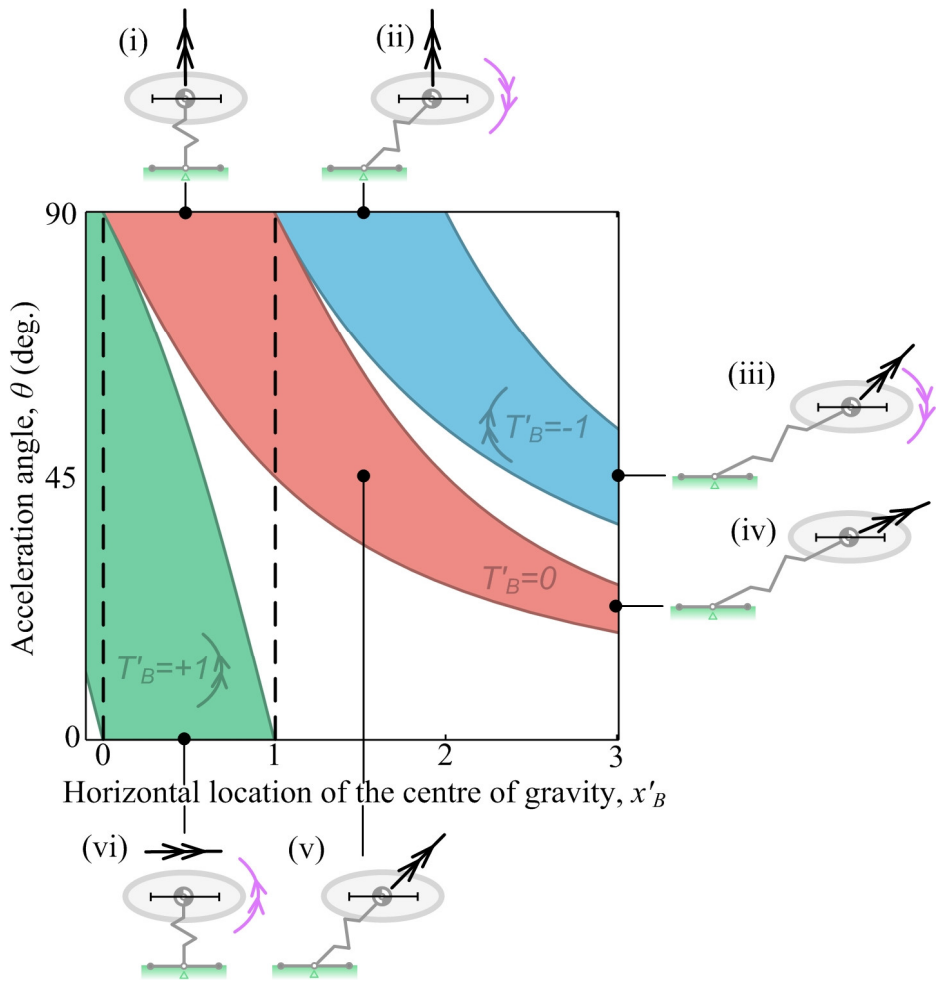

Figure 5 Stability map showing bounds on centre of gravity fore-aft location to prevent tipping. The horizontal axis represents the horizontal location of the centre of gravity, and black dashed lines mark the position of the rear-toe ($x'_b=0$) and front-toe ($x'_b=1$). The vertical axis represents the acceleration angle, 0° being horizontal acceleration and 90° being vertical. The three different coloured regions represent stability boundaries under different conditions of applied torque: red is no torque, blue is a nose-down torque and green is nose-up. Thumbnail images (i)-(vi) are example acceleration conditions with the corresponding stable centre of gravity ranges shown as black horizontal bars, with the centre of gravity illustrated at the centre of each bar; all of the cases shown are stable, with the centre of pressure located at the middle of the foot.

With no applied torque (red region in Figure 5) shallow take-off angles close to zero would require the centre of gravity to be located at an unattainably large distance ahead of the front-toe to avoid tipping backwards around the rear-toe[4]. To overcome this problem a nose-up torque (green region in Figure 5) can be used to shift the stable centre of gravity range aft to within the foot region (vi). A nose-down torque (blue region in Figure 5) has the opposite effect of shifting the required centre of gravity range forwards (Figure 5ii and iii) which creates problems for static balance in the crouch position.

Illustrations of tipping instability throughout the acceleration phase for different body acceleration conditions are shown in Figure 6 (and supplementary videos S3-S5); note that the blue arrows indicate the trajectory that the body follows, while the black arrows show net acceleration, which includes gravitational acceleration. Low body acceleration magnitudes ($\|\ddot{\mathbf{r}}_B\| < \|\mathbf{g}\|$) are chosen deliberately to illustrate the sensitivity of the CoP position to the acceleration angle.

---

[4] It should also be noted that for shallow take-off angles slipping may be more of a concern than tipping. To prevent slipping the foot-ground friction must be equal to the horizontal inertial force due to acceleration of the body. When slipping is a problem, additional vertical acceleration can be used to increase the normal component of ground reaction force, and thus increase friction.

In figure 6a (electronic supplementary material, video S3) the body accelerates horizontally, and this acceleration shifts the CoP anteriorly from its static position beneath the centre of gravity. The shift enables the system to remain stable as the centre of gravity passes forwards of the front-toe (where the equivalent static system would tip), up until the point where the CoP passes the front-toe and the system becomes dynamically unstable and tips. In figure 6b (electronic supplementary material, video S4) the body accelerates along a trajectory inclined at 45° to the horizontal. This inclination reduces the second term in equation 4, moving the CoP further forwards (and the stable centre of gravity range aft), leading to an earlier tipping point. The addition of a nose-up pitching moment in figure 6c (electronic supplementary material, video S5) increases the third term in equation 4, moving the CoP further forwards still.

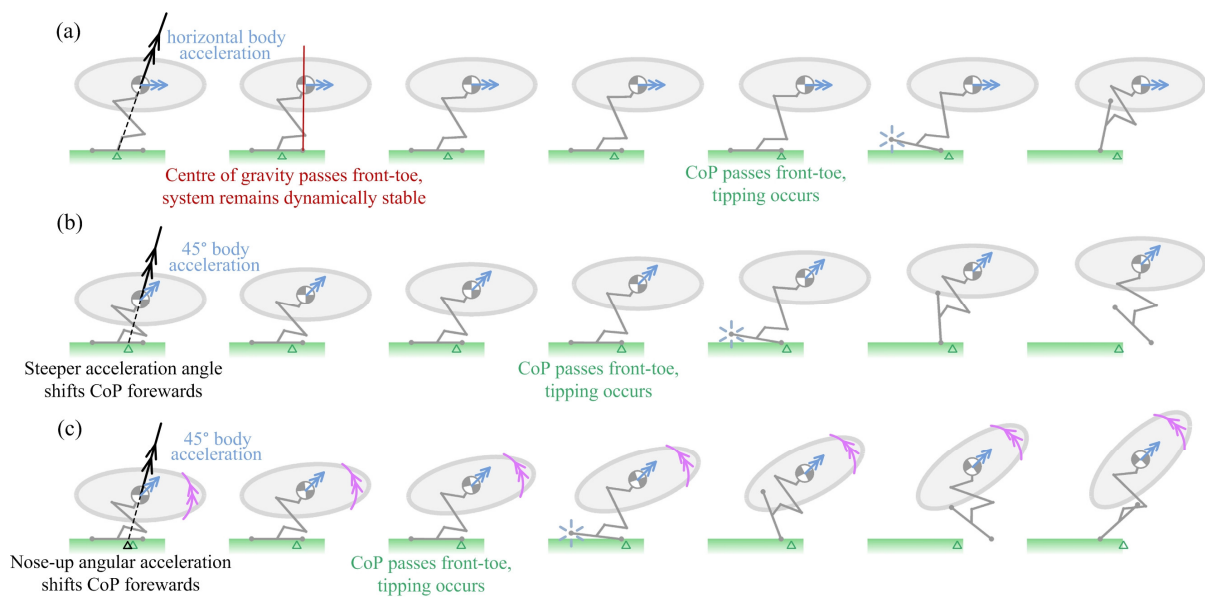

Figure 6 Simulated tipping of the 2D planar jumping system for three different acceleration conditions, with green triangles showing the moving CoP throughout the jump. See also electronic supplementary material, videos S3-S5. (a) Linear horizontal acceleration. (b) Linear acceleration, angle of 45°. (c) Linear acceleration, angle of 45° and positive (anticlockwise) angular acceleration.

## 3 Segmented leg-body model

We now introduce a realistic geometric model of the avian forelimb, body and hindlimb, Figure 7. Our objective is to develop a parsimonious model with just sufficient degrees of freedom to adequately capture the take-off dynamics from a stability point of view. The leg and body motion is planar (two dimensional) thus we assume that both legs act as a single entity. The model is designed to predict ground reactions at the foot for given input limb kinematics. For perched take-off this allows characterisation of the gripping force and torque required to achieve a given body trajectory; for ground take-off it allows evaluation of stability and body trajectory.

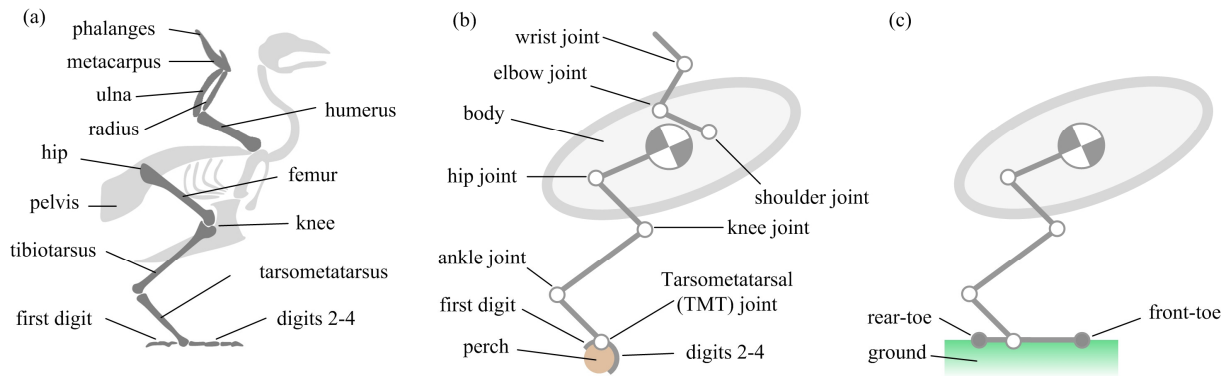

Figure 7(a) Skeletal arrangement of the avian leg, body and wing. (b,c) Abstracted multibody simulation model comprised of rigid segments connected by actuated 1 degree of freedom revolute joints at the hip, knee, ankle, tarsometatarsal (TMT) joint, and elbow, a 2dof universal joint wrist, and a 3dof spherical joint at the shoulder. (b) The 'perch model' constrains the functional foot (digits) to remain stationary on the perch. (c) The 'ground model' uses two contact points (filled circles) positioned at the rear- and front-toe to apply ground reaction forces when in contact with the ground. In the perch model (Figure 7b) it is assumed that the functional foot can provide sufficient friction to prevent rotation around the perch, so the foot remains stationary throughout the jump. The perch model captures the countermovement and acceleration dynamics up until the time where the digits begin to disengage from the perch just prior to take-off.

In the ground-model (Figure 7c) the toe is estimated as a single rigid member pivoted at the TMT joint. Discrete contact points are modelled at the front- and rear-toe. Clearly birds have many more degrees of freedom in their toes, allowing the foot to flex in response to load. One consequence of not including foot flexibility is that the contact point may switch rapidly between the heel and toe and induce a high frequency but low amplitude rocking motion. Where the contact point may vary continuously in a flexible foot the profile of ground reaction forces will be different. A second consequence is that the model cannot capture rotation about the interphalangeal joints that may occur just prior to toe-off; the rotation around a given joint would occur when the CoP passes that joint, and is therefore analogous to the foot tipping as the CoP passes the front-toe contact point. Finally, note that the ground itself is modelled as a compliant surface. This again has some effect on the dynamics of the foot, however as will be shown in the results section, ground compliance on its own does not significantly alter the overall jump performance.

In the ground-model (Figure 7c) when both the forward and rear toes are in contact with the ground the foot can react both forces and moments. In this state the leg has potentially 4 degrees of freedom (the four joint angles) and is equivalent to the perch model. When the foot rotates and contact reduces to a single point only forces can be reacted, with moments causing rotation about the contact point. In this situation the leg mechanism has 5 degrees of freedom. Modelling the rotation of the foot is significant as it changes the orientation of the entire system with respect to the ground, and hence dictates the jump trajectory. The foot orientation is not documented in the experimental literature and is therefore a core predictive output from the model that is necessary to simulate the jump. With no ground contact the airborne leg-body system can translate horizontally and vertically, and the system has a total of 7 degrees of freedom. In the perch-model (Figure 7b) the toes are kinematically constrained to always remain in contact with the perch.

It has been documented that during take-off the wing is raised during the acceleration phase such that it starts the first down stroke at or just after toe-off (4,5,44–46) and that the dynamics are driven predominantly by the hindlimbs (1). To examine the effect of the wings on stability we include wing inertial forces using a tri-segmented forelimb with actuated joints at the shoulder (3 degrees of freedom), elbow (1 degree of freedom) and wrist (2 degrees of freedom). Aerodynamic forces on the wing are calculated using the blade-element aerodynamic model presented in (46)[5]. A multibody computational model of the system was constructed with rigid-body leg segments and revolute joints, using MapleSim 6.2 (see electronic supplementary material). The control inputs to the model were time histories of the relative angles between segments. These joint angles were input into the software's inbuilt motion actuators to derive the necessary torque to drive each joint with the given kinematics. An equivalent model was trialled using tuned proportional-derivative controllers for each joint with the joint angles as setpoints; both models produced near-identical jump trajectories, with differences in take-off trajectory and speed of <0.1° and <0.2%, respectively.

In the ground model normal ground reaction forces at the toe contact points were modelled as unilateral critically damped springs, capable of applying force only in the positive y-direction. In the absence of experimental data for bird foot stiffness, the springs were assigned a nominal stiffness value of $40 kNm^{-1}$ based on the previous literature on human gait modelling (47); section 4 will illustrate that using alternative values (e.g. those used in robotic simulation (48)) has a limited impact on the overall jump trajectory, but does affect ground reaction force time histories. Frictional forces were imposed to prevent sliding using an overdamped bilateral actuator at each contact point with a damping coefficient of $10^3$ $Nsm^{-1}$, which constrained lateral displacements to <1% of the toe length. In the perch model, the toes are kinematically constrained to remain on the perch.

The system physical properties were modelled on the guinea fowl (*Numida meleagris*) as the most detailed quantitative measurements of avian leg segment kinematics in jumping are available for this species (2), and on the diamond dove (*Geopelia cuneata*) as high fidelity recordings of kinematics are available for both the leg and wing (49). Using these two species also allows us to interrogate the jumping dynamics at differing scales. Leg joint angle kinematics were obtained for the guinea fowl by manually extracting 15 digitised data points of angular position time histories from (2) and interpolating these through the jumping time period using cubic spline interpolation; this gives finite angular acceleration at the joints, which varies piecewise linearly with time. Leg joint angle kinematics were obtained for the diamond dove by manually overlaying linear segments onto digitised frames in (49); segment angular positions were measured at 10 equally space time points throughout the cycle, and interpolated using a cubic spline. Wing kinematics for the diamond dove were defined by estimating the upstroke period and wing elevation amplitude from digitised frames in (49); these are then applied using a previous sinusoidal model of wing extension and elevation (46), with the assumption that the wing is initially fully flexed and extends to its maximum length at the end of the upstroke.

For the guinea fowl the total system mass was 1.42kg (2) and leg segment lengths were taken from (50), while for the diamond dove the mass was 0.052kg (49) and the leg segments lengths were scaled from digitised images using a previously recorded tarsometatarsal length (33). The body moment of inertia was defined by modelling it as a prolate spheroid (51) with cross sectional area defined from an allometric model (52), and an approximated slenderness (length/width) ratio of 2. The wing mass was defined using an allometric model (53), and the mass was distributed equally across each of the 3 forelimb segments, with the centre of mass approximated to lie at the geometric centre of each segment.

All simulations were setup using a 4th order Runge Kutta numerical integration scheme with a fixed timestep size of $10^{-5}$ s. A variable step numerical integration scheme can be employed to reduce computational cost for an equivalent degree of accuracy, but for simplicity of examining numerical convergence a fixed-timestep solver was used in this study. Halving the timestep size yielded <1% change in take-off linear and angular velocity; <0.5% change was seen in ground reaction force, including during high frequency dynamic events where the foot-ground contacts were made and broken. A sensitivity analysis of the inertial input parameters was conducted using the guinea fowl model (see supplementary material 2) and found the jump dynamics to be relatively insensitive to the system moment of inertia: doubling the moment of inertia (assuming a body slenderness ratio of 3) increased the take-off trajectory by 6° and speed by <1%. Moreover, only a 5° reduction in take-off trajectory and <1% reduction in speed are seen when comparing a model with realistic mass and inertia properties for all leg segments (taken from (54)) to one with massless legs; the massless leg model is used in subsequent simulations as it is conceptually simpler to describe using the analytical tools developed in section 2. Finally, a fore-aft shift (+/- *x*-direction) in position of the body centre of gravity by a distance of 10% of the functional foot length led to a change in take-off trajectory of 2° and take-off speed of <1%.

---

[5] The blade-element method approximates the 3d wing as a series of 2d aerofoils. Local wind conditions for each aerofoil are derived from wing kinematics and the induced flow field. Previous aerodynamic data on avian aerofoils are then used with the local wind conditions to determine the forces and moments on each aerofoil, before summing them for the entire wing.

## 4. Results and Discussion

### 4.1 Jumping dynamics and stability

The perch model illustrates the countermovement, followed by rapid extension of the leg and a concurrent extension and elevation of the wing (Figure 8a; electronic supplementary, video S6). Examining the dynamics of this manouver finds a positive (forwards) initial horizontal acceleration of the body, or equivalently a positive horizontal ground reaction force, which diminishes over time (Figure 9a). The horizontal ground reaction force becomes negative at $t/T_{t\text{-off}} \sim 0.7$, which is not believed to be representative of the biological system, and occurs here due to the model neglecting the disengagement of the toes from the perch (see section 3); later it will be shown that the ground model predicts the reaction force reasonably closely to the experimental measurements when the foot is fully engaged with the ground. For this reason the simulation results for $t/T_{t\text{-off}} > 0.7$ are not considered in the following discussions of jumping dynamics or stability, and as such are greyed out in Figure 9, but included here for completeness.

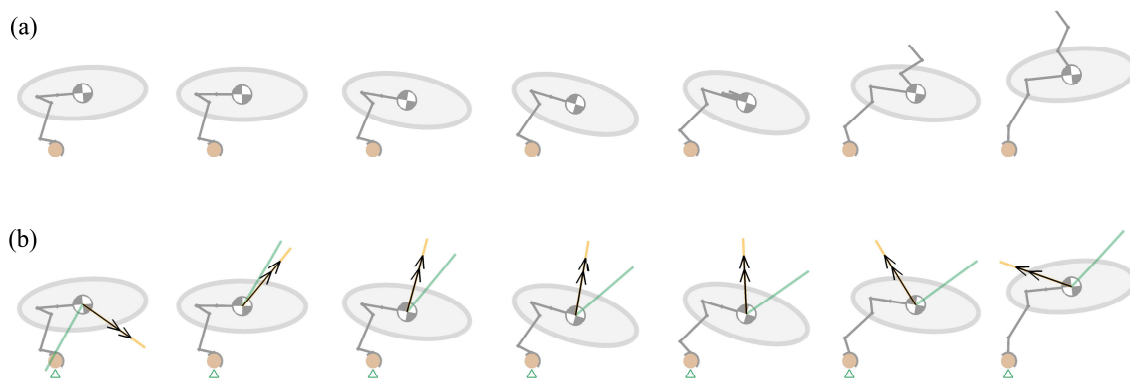

Figure 8 Frames taken from a simulated jumping take-off of a diamond dove (*Geopelia cuneata*) at time intervals of 0.033s, using leg joint angle kinematics obtained from digitised images from experimental recordings (49). The sequence begins at the initiation of the countermovement (as defined in (33)). The jump sequence is shown over the time period of experimental data capture in (49). (a) shows the frames incorporating the wing model that uses a sinusoidal elevation and extension kinematics model in (46), with the amplitude and phase estimated from videos in (49). (b) shows the net acceleration vector (black arrow), centre of pressure (green triangle), stability window (yellow line) and zero-torque stability window (green line)

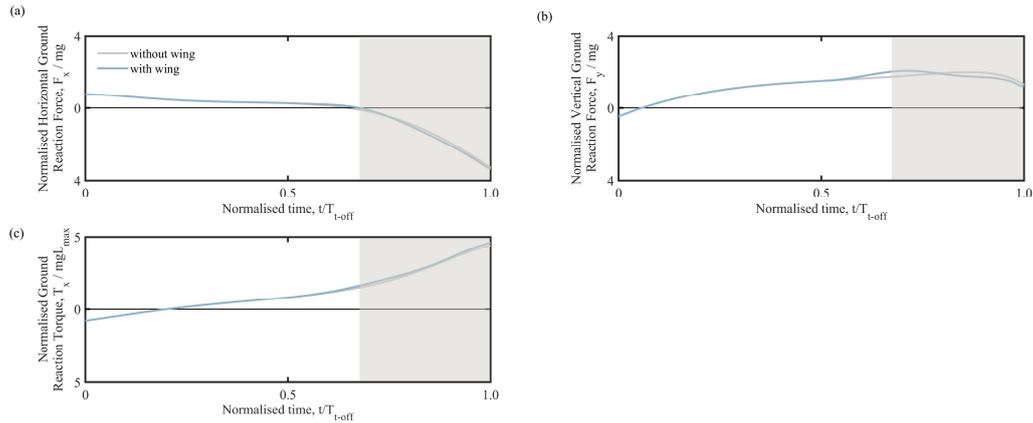

Figure 9 Time histories of ground reactions from simulations of a jumping diamond dove, *Geopelia cuneata* (49). Results are plotted against time normalised by the toe-off time, $T_{t\text{-off}}$, which is defined here as the time in the experiments at which there is complete loss of contact between the toe and the perch, assessed through visual inspection of videos. Simulations are shown for the cases with the wing (blue line) and without the wing (gray line). Ground reactions torque (c) is normalised the product of the body weight and the maximum leg length, $L_{max}$, given as the sum of the four leg segment lengths. Positive torque is nose-up.

The vertical ground reaction force is initially negative, meaning that the downward acceleration magnitude exceeds that of freefall as the gripping foot pulls the body downwards towards the perch. (Figure 9b). The ground reaction force increases to zero at $t/T_{t\text{-off}} \sim 0.07$, and continues to increase to a peak of $F_y/mg \sim 2$. This body acceleration is modest in comparison to previously recorded vertical jumping accelerations of over $5g$ in guinea fowl (2), $4g$ in starlings (1) and almost $8g$ in quails (1), but is plausible for diamond doves which has previously been recorded to have take-off accelerations of 2-$3g$ (33).

The ground reaction torque is negative at the start of the countermovement (Figure 9c), meaning that gripping prevents the foot rotating backwards (anticlockwise) around the perch. The ground reaction torque increases over time and becomes positive, where gripping then prevents forward (clockwise) rotating of the foot around the perch. The normalised torque reaches a value of $\sim 1.5$ before the toes begin to disengage: the equivalent torque would be felt by a stationary diamond dove if its centre of mass were displaced ahead of its toes a distance of 1.5 times the length of the fully extended leg. To the best of the authors' knowledge this is the first quantitative estimation of ground reaction torques in avian jumping.

The wing contributes very little to the mean ground reaction forces and torques for the most part of the take-off and it may be ignored from a stability point of view. The greatest influence of the wing on the system dynamics is that it increases the peak vertical ground reaction force by around 15% just prior to the toes disengaging (Figure 9b). At this time, the ground reaction torques increase by ~9% if the wing is included. The magnitudes of these predicted forces and torques contribute to the body of evidence suggesting that the wing has a secondary influence on the dynamics of avian jumping (1,2,33). Inclusion of wing aerodynamic forces has a negligible effect on ground reaction forces and torques (< 1%) and so are omitted from Figure 9 for clarity.

Figure 8b shows that the direction of the net acceleration, $\mathbf{R}_p/m$, (black arrows) is vectored downwards and forwards at the start of the countermovement, and then upwards and forwards through the remainder of the countermovement and the leg extension phase. Note that in the 5th-8th frames in figure 8b the net acceleration vectors point backwards unrealistically, again due to the omission in the model of the toes disengaging from the perch. The centre of pressure (green triangle) remains directly beneath the single contact point at the TMT joint, as required by the imposed constraint in the model that prevents the foot rotating. The net acceleration vector and its stability window (yellow sector) are aligned by definition due to the foot gripping. With lower gripping torque the foot would rotate around the perch, the centre of pressure would be displaced from beneath the ground contact point, and the net acceleration vector would not be aligned with the stability window.

The green stability window for the zero-torque system, diverges from the yellow stability window through the leg extension phase. This is due to the applied gripping torque on the perch becoming increasingly dominant in preventing rotation of the foot. Without this applied torque the system could only remain stable by using a shallower acceleration angle to align the acceleration vector with the green window, which also aligns with the TMT contact point. Figure 8b frames 4-5 depict the time at which the real bird disengages its toes from the perch. At this time the stable jump angle (yellow window) is close to vertical, and the gripping torque extends the take-off angle by an additional 45° compared to a case with no gripping capability, as shown by the angle between the yellow and green stability windows.

While a complete analysis of dynamic similarity is beyond the scope of this work, a preliminary scaling analysis was conducted by geometrically scaling the diamond dove model but retaining the same joint kinematics. A geometric increase in scale leads to an increase in magnitude of normalised ground reaction forces and torques; larger birds require greater gripping torque relative to their mass and leg length in order to maintain tipping stability. If body mass alone is increased with fixed leg geometry and kinematics, the normalised reaction forces and torques remain constant.

The ground-model captures the aggregate jumping manouver seen experimentally in the guinea fowl, with the leg driving both horizontal and vertical body displacement combined with a nose-up rotation (Figure 10a, b; electronic supplementary material, videos S7, S8). The take-off performance, characterised by the take-off velocity and orientation of the body, is relatively insensitive to large changes in ground stiffness seen between the 'soft' and 'hard' ground models. The system stability just prior to take-off (5$^{th}$ frame) is also similar between the models, as both tip onto the front-toe prior to take-off with <2° difference in the foot orientation. However the contact stiffness does influence the detailed high frequency stability dynamics near the beginning of the jump, as evidenced in frames 1-3 by the differences in widths of the stability windows and differences in location of the CoP.

In the soft ground model the CoP is initially near the centre of the foot region (first 3 frames of Figure 10a). Correspondingly, the acceleration vector (black arrow) is within the stability window (yellow sector), and both front- and rear-toe remain in contact with the ground. The CoP then moves just ahead of the front-toe as the rear-toe leaves the ground (4$^{th}$ frame), and the stability window sector collapses to a line, as expected. Throughout the remainder of the acceleration phase the acceleration vector is in near-alignment with the stability window, and the CoP is beyond but close to the front-toe, highlighting that the system is close to being neutrally stable.

In the firm ground model the CoP moves from front-, to rear-, to front-toe as the bird rocks forwards and backwards (first 3 frames of Figure 10b). If the leg were to remain static with the initial posture shown, the system would be statically stable, so the rocking observed is a dynamic effect due to kinematics of the leg joints. The next critical event in the sequence is a stutter (4$^{th}$ frame of Figure 10b), where both front-and rear-contact points on the toe temporarily leave the ground; the CoP symbol is ommitted as the ground reaction force is zero. The foot then regains contact with the ground, before tipping forwards as the CoP passes the front-toe, similar to the soft-ground model.

For both the soft- and firm-ground models this toe-tip immediately preceeds take-off and the beginning of the flight phase, with the firm-ground model taking off earlier. Within flight the leg continues to extend, akin to the example presented in Figure 3b, exemplifying again that the toe-off condition does not correspond to the leg becoming fully extended.

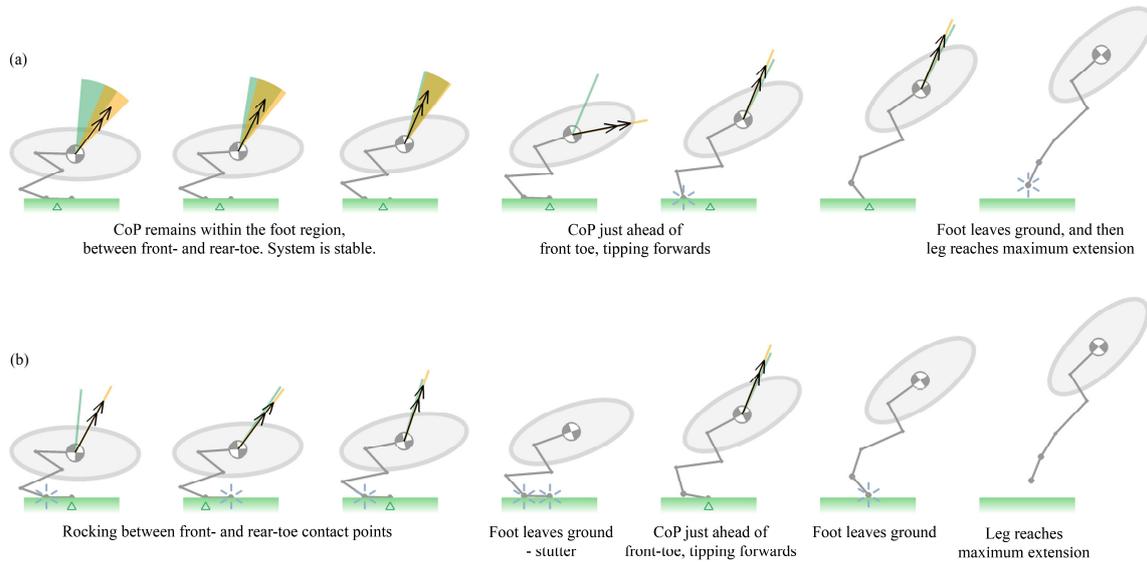

Figure 10 Simulated jumping take-off of a guinea fowl, *Numida meleagris*, using the ground-model. Frames taken at time intervals of 0.02s, using experimentally recorded joint angle kinematics (2). The sequence begins in the crouched posture, where the body centre of gravity is recorded as being at its lowest point during the preparation to jump (2). The leg is modelled as being massless, with the total mass of the bird (1.42kg) being concentrated at the body centre of gravity. The body moment of inertia is predicted by modelling the body as an ellipsoid with a slenderness ratio of 2. (a) Uses a 'soft' ground contact stiffness of $2kNm^{-1}$, taken from previous models of robotic terrestrial locomotion (48). (b) Uses a 'firm' ground contact stiffness of $40kNm^{-1}$ taken from previous simulations of human walking (47).

Examining the jumping dynamics in more detail reveals that the model simulates the trends in body linear displacements reasonably closely to the experimental values (Figure 11a, c). The simulated vertical impulse (Figure 11d) follows the experimental trends during the initial jump phase up until the time at which stutter occurs for the firm- and hard-ground models. Stutter is characterised by the plateau in impulse and by the acceleration falling to $-g$ at $t/T_{\text{t-off}} \sim 0.5$ (Figure 11f), which is narrowly avoided by the soft-ground model. Following stutter, the regaining of ground contact leads to a rapid increase in impulse, with final impulse at toe-off predicted to be around 20% larger than the experimental result. However, the acceleration predictions (and equivalently, force) are found to be more sensitive to the ground stiffness than the integrated quantities of velocity, impulse, or displacement, and show the greatest difference from experimental measurements (Figure 11b).

For firm- and hard- ground models that predict a stutter, the post-stutter touchdown yields an unrealistically high prediction of ground reaction force ($F_y/mg > 20$ at $t/T_{\text{t-off}} \sim 0.6$). The two earlier peaks in ground reaction force are concurrent with the rocking between front- and rear-toe contact points seen in Figure 10b. These peaks in force are partly an artefact of using a rigid linear foot segment with only front- and rear- ground contact points, and would be alleviated in a foot model with a greater number of articulated joints and contact points. The additional compliance in the soft-ground model eliminates this ground contact intermittency and makes a reasonable prediction of ground reaction force until $t/T_{t\text{-off}} \sim 0.4$; after this the simulated forces deviate from the experiments as the heel loses contact with the ground.

The rocking and stuttering dynamics are not seen in avian jumping take-offs, but are physical consequences of the measured kinematics rather than numerical artefacts from simulation. Both are eliminated in the simulation by increasing the compliance of the contact model. Equivalently a more realistic skeletal model that included joint compliance would also reduce these phenomena and increase accuracy in predicting ground reaction forces. Preliminary trials were conducted with an additional joint and segment modelled in the toe, however these models still stuttered during the jump. For more accurate prediction of ground reaction force profiles future models would need to address this issue of compliance. However, for predicting take-off speed and trajectory the model is relatively insensitive to the contact model stiffness.

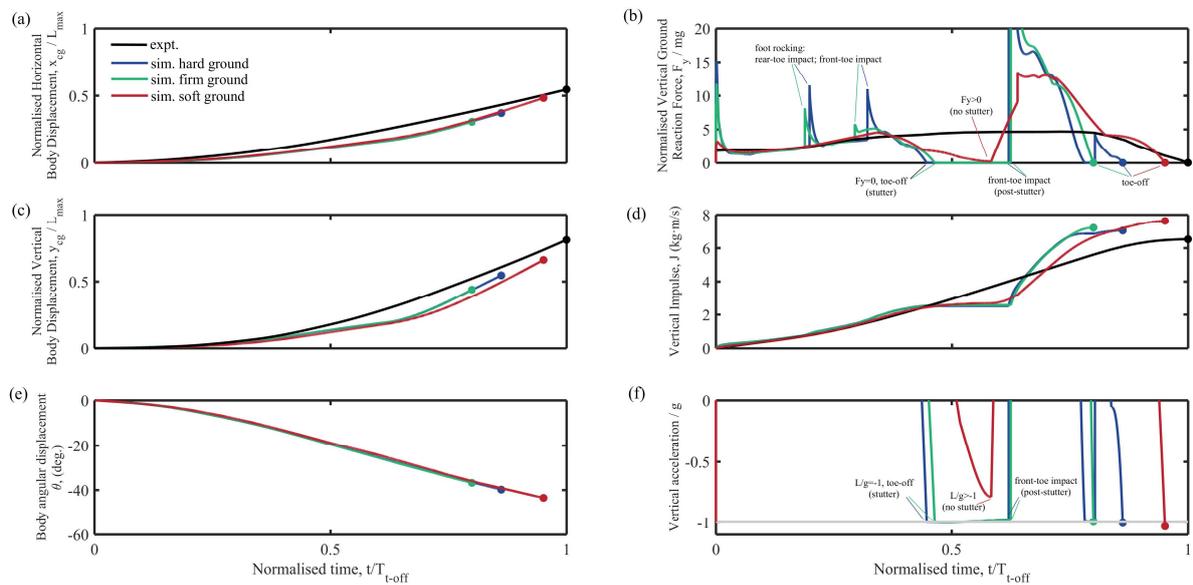

Figure 11 Time histories of body kinematics and dynamics from simulations (blue, green and red) and experimental measurements (black) of a jumping guinea fowl, *Numida meleagris*. Results are plotted against time normalised by the toe-off time for the experimental measurements, $T_{t\text{-off}}$; circular symbols highlight the toe-off time for each case. Simulations are shown for three different ground contact stiffnesses, with the 'soft' and 'hard' ground contact stiffnesses of $2kNm^{-1}$, and $80kNm^{-1}$, respectively, taken from previous models of robotic terrestrial locomotion (48), and the 'firm' stiffness of $40kNm^{-1}$ taken from previous simulations of human walking (47). Body linear and angular displacements are plotted relative to the position in the crouch posture as defined by (2). Linear displacements are normalised against the maximum leg length, $L_{max}$, given as the sum of four leg segment lengths. See also electronic supplementary material, videos S7 and S8. Light grey horizontal line in (f) represents the limiting case of vertical acceleration equal to $-g$.

## 4.2 Modulating toe-off and tipping

As a final point of investigation this section summarises how the joint angle kinematics of the guinea fowl model can be strategically modified in order to adjust the dynamic events of toe-off and tipping. The ambition is to serve as a precursor to future investigations into methods of control used in avian jumping. Full details of the analysis are included in supplementary material 4.

During the final extension phase of the jump increasing the angular rate of the TMT joint above the experimental values causes a delay in toe-off (supplementary material figure S4). This adjustment to the TMT joint kinematics delays toe-off by delaying the time at which the body acceleration falls to $-g$ (see theory section 2.3). In the preceding phase of the jump prior to the leg extending there is a characteristic drop in TMT joint angle (2). Modifying the kinematics to create an even greater drop in TMT angle causes the CoP to shift backwards, delaying the time at which the system tips onto the front-toe (supplementary material Figure S5).

## 5. Conclusions

This paper complements previous experimental investigations into avian take-off by examining the stability constraints that jumping birds must work within, with the notion that stability is a prerequisite for high performance. Analytical models using CoP location as a stability metric provided a novel graphical interpretation of the range of centres of gravity and the 'windows' of body acceleration angle for which stability can be maintained. A shallow-angled forward jump commonly used for initiating wing-borne flight requires the centre of gravity to be well forward of the foot to prevent rear-toe tipping during acceleration. Inclusion of an additional head up angular acceleration brings the required centre of gravity range aft and thus potentially simplifies the kinematics required for a successful take-off.

The multibody simulation of a diamond dove in a prehensile (perch-gripping) jump predicts that the aerodynamic force is negligible during the wing upstroke, and that the wing inertial effects contribute very little to the mean ground reaction force and torque. The model also provides the first quantification of the foot-gripping couple required to maintain stability: the magnitude of this moment is equivalent to the weight of the bird acting at a moment arm approximately 1.5 leg lengths ahead of the perch. The prehensile jump allows for greater flexibility in take-off by extending the range of jump angles for the diamond dove by around 45°.

Simulations of a guinea fowl performing a ground take-off showed the centre of pressure passing the front of the toe, leading to toe-tipping immediately prior to take-off, which is also seen in the experiments. The system is dynamically stable through the initial acceleration phase, and then close to neutrally stable at toe-off. This is evidenced by the proximity of the centre of pressure to the front toe and also the near alignment of the net acceleration and the stability window. Ground-based jumps can utilise less reacted torque from the ground compared to prehensile jumps. For prehensile jumps the gripping torque peaks just prior to toe-off.

With a firm ground model the simulation exhibits a stuttering behaviour with rapid loss and gain of foot contact. This can be reduced by increasing the compliance of the ground contact model, or equivalently by introducing joint compliance, something that has been neglected in this model. Stuttering is not a simulation artefact, but a physical consequence of the experimental input kinematics giving a non-uniform acceleration profile that induces strong intermittency in the ground contact forces. The simulated stuttering illustrates a limitation of using a kinematically driven model with rigid joints. Despite this, the finding is still biologically relevant in illustrating the physical conditions that real birds must avoid in order to prevent premature take-off or unwanted high frequency dynamics that may affect mechanical control; these same physical conditions are targeted by animals that do perform deliberate stutter jumps.

Within the limitations of the model there is a reasonably good agreement for the integrated quantities of velocity and displacement between the experimental measurements and simulation. Simulated ground reaction force is less well captured, and the firm ground model has been shown to introduce high frequency variations in the predicted force, which are not seen in the experiments. These force fluctuations arise from stuttering, and also from the model rocking between the front and rear contact points. Despite these, the bulk simulation output of take-off speed and trajectory is seen to be relatively insensitive to input physical parameters, including ground contact stiffness. As such the multibody model is regarded as robust tool for predicting take-off performance, and complements existing bird flight performance models that focus on the steady cruising flight conditions.

The verified guinea fowl simulation model was used to evaluate the behaviours associated with stuttering and tipping in a representative jumping bird. It was shown that by adjusting the TMT angle kinematics the vertical acceleration profile can be modified, causing a delay or promotion of toe-off that is consistent with the analytical predictions. Tipping instability was investigated in a similar fashion to invoke both stable and unstable conditions prior to toe-off, demonstrating that the TMT joint can be used to provide a degree of low-level control in avian jumping take-off.

### Data Accessibility

Our data are included in the electronic supplementary material.

### Competing Interests

We have no competing interests.

### Authors' Contributions

B.P., G.S. and W.C. conceived the original study and helped draft the manuscript. B.P. and G.S. developed the computational multibody models, and B.P. performed the simulations and conducted the analysis. B.P and W.C. developed the analytical models. All authors gave final approval for publication.

### Funding

No funding supported this research.

### Animal Ethics

We were not required to complete an ethical assessment prior to conducting this research.

### Research Ethics

We were not required to complete an ethical assessment prior to conducting this research.

### Permission to carry out fieldwork

No permissions were required prior to conducting this research.

### Acknowledgements

We would like to thank William Sellers for his valuable suggestions on the style and content of the manuscript, particularly the introduction.

## References

1. Earls KD. Kinematics and mechanics of ground take-off in the starling Sturnis vulgaris and the quail Coturnix coturnix. J Exp Biol. 2000 Feb;203(Pt 4):725–39.

2. Henry HT, Ellerby DJ, Marsh RL. Performance of guinea fowl Numida meleagris during jumping requires storage and release of elastic energy. J Exp Biol. 2005 Sep 1;208(17):3293–302.

3. Tobalske BW, Altshuler DL, Powers DR. Take-off mechanics in hummingbirds (Trochilidae). J Exp Biol. 2004 Mar 15;207(8):1345–52.

4. Heppner FH, Anderson JGT. Leg Thrust Important in Flight Take-Off in the Pigeon. J Exp Biol. 1985 Jan 1;114(1):285–8.

5. Bonser R, Rayner J. Measuring leg thrust forces in the common starling. J Exp Biol. 1996 Feb 1;199(2):435–9.

6. Rayner JMV. A New Approach to Animal Flight Mechanics. J Exp Biol. 1979 Jun 1;80(1):17–54.

7. Nudds RL, Bryant DM. The energetic cost of short flights in birds. J Exp Biol. 2000 May;203(Pt 10):1561–72.

8. Pandy MG, Zajac FE, Sim E, Levine WS. An optimal control model for maximum-height human jumping. J Biomech. 1990;23(12):1185–98.

9. Bobbert MF, Gerritsen KGM, Litjens MCA, Van Soest AJ. Why is countermovement jump height greater than squat jump height? Med Sci Sports Exerc. 1996 Nov;28(11):1402–1412.

10. Bobbert MF. Dependence of human squat jump performance on the series elastic compliance of the triceps surae: a simulation study. J Exp Biol. 2001 Feb;204(Pt 3):533–42.

11. Alexander RM. Leg design and jumping technique for humans, other vertebrates and insects. Philos Trans R Soc Lond B Biol Sci. 1995 Feb 28;347(1321):235–48.

12. Sellers WI. A biomechanical investigation into the absence of leaping in the locomotor repertoire of the slender loris (Loris tardigradus). Folia Primatol Int J Primatol. 1996;67(1):1–14.

13. Kargo WJ, Nelson F, Rome LC. Jumping in frogs: assessing the design of the skeletal system by anatomically realistic modeling and forward dynamic simulation. J Exp Biol. 2002 Jun 15;205(12):1683–702.

14. Bennet-Clark HC, Lucey ECA. The Jump of the Flea: A Study of the Energetics and a Model of the Mechanism. J Exp Biol. 1967 Aug 1;47(1):59–76.

15. Armour R, Paskins K, Bowyer A, Vincent J, Megill W, Bomphrey R. Jumping robots: a biomimetic solution to locomotion across rough terrain. Bioinspir Biomim. 2007 Sep;2(3):S65-82.

16. Niiyama R, Nagakubo A, Kuniyoshi Y. Mowgli: A Bipedal Jumping and Landing Robot with an Artificial Musculoskeletal System. In: 2007 IEEE International Conference on Robotics and Automation. 2007. p. 2546–51.

17. Haldane DW, Plecnik MM, Yim JK, Fearing RS. Robotic vertical jumping agility via series-elastic power modulation. Sci Robot. 2016 Dec 6;1(1):eaag2048.

18. Cofer D, Cymbalyuk G, Heitler WJ, Edwards DH. Control of tumbling during the locust jump. J Exp Biol. 2010 Oct 1;213(19):3378–87.

19. Gvirsman O, Kosa G, Ayali A. Dynamics and stability of directional jumps in the desert locust. PeerJ. 2016 Sep 28;4:e2481.

20. Vukobratovic M, Frank AA, Juricic D. On the Stability of Biped Locomotion. IEEE Trans Biomed Eng. 1970 Jan;BME-17(1):25–36.

21. Vukobratović M, Borovac B. Zero-moment point — thirty five years of its life. Int J Humanoid Robot. 2004 Mar 1;01(01):157–73.

22. Vukobratović M, Stepanenko J. On the stability of anthropomorphic systems. Math Biosci. 1972 Oct 1;15(1):1–37.

23. Shih C-L. The Dynamics and Control of a Biped Walking Robot With Seven Degrees of Freedom. J Dyn Syst Meas Control. 1996 Dec 1;118(4):683–90.

24. Li Q, Takanishi A, Kato I. A biped walking robot having a ZMP measurement system using universal force-moment sensors. In: Proceedings IROS '91:IEEE/RSJ International Workshop on Intelligent Robots and Systems '91. 1991. p. 1568–73 vol.3.

25. Ugurlu B, Kawamura A. Bipedal Trajectory Generation Based on Combining Inertial Forces and Intrinsic Angular Momentum Rate Changes: Eulerian ZMP Resolution. IEEE Trans Robot. 2012 Dec;28(6):1406–15.

26. Ugurlu B, Kawamura A. ZMP-Based Online Jumping Pattern Generation for a One-Legged Robot. IEEE Trans Ind Electron. 2010 May;57(5):1701–9.

27. Sardain P, Bessonnet G. Forces acting on a biped robot. Center of pressure-zero moment point. IEEE Trans Syst Man Cybern - Part Syst Hum. 2004 Sep;34(5):630–7.

28. Andrada E, Rode C, Sutedja Y, Nyakatura JA, Blickhan R. Trunk orientation causes asymmetries in leg function in small bird terrestrial locomotion. Proc R Soc Lond B Biol Sci. 2014 Dec 22;281(1797):20141405.

29. Andrada E, Haase D, Sutedja Y, Nyakatura JA, Kilbourne BM, Denzler J, et al. Mixed gaits in small avian terrestrial locomotion. Sci Rep. 2015 Sep 3;5:13636.

30. Hof AL, Gazendam MGJ, Sinke WE. The condition for dynamic stability. J Biomech. 2005 Jan;38(1):1–8.

31. McAndrew Young PM, Wilken JM, Dingwell JB. Dynamic margins of stability during human walking in destabilizing environments. J Biomech. 2012 Apr 5;45(6):1053–9.

32. Backus SB, Sustaita D, Odhner LU, Dollar AM. Mechanical analysis of avian feet: multiarticular muscles in grasping and perching. R Soc Open Sci. 2015 Feb 1;2(2):140350.

33. Provini P, Tobalske BW, Crandell KE, Abourachid A. Transition from leg to wing forces during take-off in birds. J Exp Biol. 2012 Dec 1;215(23):4115–24.

34. Lees JJ, Folkow LP, Codd JR, Nudds RL. Seasonal differences in jump performance in the Svalbard rock ptarmigan (Lagopus muta hyperborea). Biol Open. 2014 Apr 15;3(4):233–9.

35. Demes B, Jungers WL, Fleagle JG, Wunderlich RE, Richmond BG, Lemelin P. Body size and leaping kinematics in Malagasy vertical clingers and leapers. J Hum Evol. 1996 Oct 1;31(4):367–88.

36. Günther MM, Ishida H, Kumakura H, Nakano Y. The jump as a fast mode of locomotion in arboreal and terrestrial biotopes. Z Morphol Anthropol. 1991;78(3):341–72.

37. Essner RL. Three-dimensional launch kinematics in leaping, parachuting and gliding squirrels. J Exp Biol. 2002 Aug;205(Pt 16):2469–77.

38. Aguilar J, Lesov A, Wiesenfeld K, Goldman DI. Lift-Off Dynamics in a Simple Jumping Robot. Phys Rev Lett. 2012 Oct 26;109(17):174301.

39. Aguilar J, Goldman DI. Robophysical study of jumping dynamics on granular media. Nat Phys. 2015 Nov 30;12(3):nphys3568.

40. Hale E, Schara N, Burdick J, Fiorini P. A minimally actuated hopping rover for exploration of celestial bodies. In: Proceedings 2000 ICRA Millennium Conference IEEE International Conference on Robotics and Automation Symposia Proceedings (Cat No00CH37065). 2000. p. 420–7 vol.1.

41. Goswami A. Foot rotation indicator (FRI) point: a new gait planning tool to evaluate postural stability of biped robots. In: Proceedings 1999 IEEE International Conference on Robotics and Automation (Cat No99CH36288C). 1999. p. 47–52 vol.1.

42. Tajima R, Honda D, Suga K. Fast running experiments involving a humanoid robot. In: 2009 IEEE International Conference on Robotics and Automation. 2009. p. 1571–6.

43. Simpson SF. The flight mechanism of the pigeon Columbia livia during take-off. J Zool. 1983 Jul 1;200(3):435–43.

44. Dial KP, Biewener AA. Pectoralis Muscle Force and Power Output During Different Modes of Flight in Pigeons (columba Livia). J Exp Biol. 1993 Mar 1;176(1):31–54.

45. Gatesy SM, Dial KP. Tail Muscle Activity Patterns in Walking and Flying Pigeons (columba Livia). J Exp Biol. 1993 Mar 1;176(1):55–76.

46. Parslew B. Predicting power-optimal kinematics of avian wings. J R Soc Interface. 2015 Jan 6;12(102):20140953.

47. Gilchrist LA, Winter DA. A two-part, viscoelastic foot model for use in gait simulations. J Biomech. 1996 Jun 1;29(6):795–8.

48. Dallali H. Modelling and Dynamics Stabilisation of a Complicant Humanoid Robot, CoMan [Internet]. [Manchester, UK]: University of Manchester; 2012 [cited 2017 Nov 4]. Available from: http://oatd.org/oatd/record?record=%22oai%5C%3Aescholar.manchester.ac.uk%5C%3Auk-ac-man-scw-160712%22

49. Provini P, Abourachid A. Whole-body 3D kinematics of bird take-off: key role of the legs to propel the trunk. Sci Nat. 2018 Feb 1;105(1–2):12.

50. Gatesy SM, Biewener AA. Bipedal locomotion: effects of speed, size and limb posture in birds and humans. J Zool. 1991 May 1;224(1):127–47.

51. Norberg RÅ. Treecreeper Climbing; Mechanics, Energetics, and Structural Adaptations. Ornis Scand Scand J Ornithol. 1986;17(3):191–209.

52. Hedenström A, Rosén M. Body Frontal Area in Passerine Birds. J Avian Biol. 2003;34(2):159–62.

53. van der Berg C, Rayner JMV. The moment of inertia of bird wings and the inertial power requirement for flapping flight. J Exp Biol. 1995;198:1655–64.

54. Rubenson J, Marsh RL. Mechanical efficiency of limb swing during walking and running in guinea fowl (Numida meleagris). J Appl Physiol Bethesda Md 1985. 2009 May;106(5):1618–30.